\documentclass[journal,transmag]{IEEEtran}
\usepackage{cite}
\usepackage{textcomp}
\ifCLASSINFOpdf
   \usepackage[pdftex]{graphicx}
   \graphicspath{{../pdf/}{../jpeg/}}
   \DeclareGraphicsExtensions{.pdf,.jpeg,.png}
\else
\fi
\usepackage{amsmath}
\usepackage{breqn}
\begin{document}
%
\title{Multimode Nonlinear Fiber Optics: Massively Parallel Numerical Solver, Tutorial and Outlook}


\author{\IEEEauthorblockN{Logan G. Wright\IEEEauthorrefmark{1},
Zachary M. Ziegler\IEEEauthorrefmark{1},
Pavel M. Lushnikov\IEEEauthorrefmark{2},
Zimu Zhu\IEEEauthorrefmark{1}, 
M. Amin Eftekhar\IEEEauthorrefmark{3}, \\
Demetrios N. Christodoulides\IEEEauthorrefmark{3}
, and Frank W. Wise\IEEEauthorrefmark{1}}
\IEEEauthorblockA{\IEEEauthorrefmark{1}School of Applied and Engineering Physics,
Cornell University, Ithaca, New York 14853, USA}
\IEEEauthorblockA{\IEEEauthorrefmark{2}Department of Mathematics and Statistics, University of New Mexico, Alberquerque, New Mexico 87131, USA}
\IEEEauthorblockA{\IEEEauthorrefmark{2}CREOL, College of Optics and Photonics,
University of Central Florida, Orlando, Florida 32816, USA}

\thanks{Manuscript received December 1, 2012; revised August 26, 2015. 
Corresponding author: L.G. Wright (email: lgw32@cornell.edu).}}

\markboth{Journal of \LaTeX\ Class Files,~Vol.~14, No.~8, August~2015}%
{Shell \MakeLowercase{\textit{et al.}}: Bare Demo of IEEEtran.cls for IEEE Transactions on Magnetics Journals}
%



\IEEEtitleabstractindextext{%
\begin{abstract}
Building on the scientific understanding and technological infrastructure of single-mode fibers, multimode fibers are being explored as a means of adding new degrees of freedom to optical technologies such as telecommunications, fiber lasers, imaging, and measurement. Here, starting from a baseline of single-mode nonlinear fiber optics, we introduce the growing topic of multimode nonlinear fiber optics. We demonstrate a new numerical solution method for the system of equations that describes nonlinear multimode propagation, the generalized multimode nonlinear Schr\"{o}dinger equation. This numerical solver is freely available, implemented in MATLAB\textsuperscript{\textregistered} and includes a number of multimode fiber analysis tools. It features a significant parallel computing speed-up on modern graphical processing units, translating to orders-of-magnitude speed-up over the split-step Fourier method. We demonstrate its use with several examples in graded- and step-index multimode fibers. Finally, we discuss several key open directions and questions, whose answers could have significant scientific and technological impact. 
\end{abstract}

}

\maketitle

\IEEEdisplaynontitleabstractindextext

%
\IEEEpeerreviewmaketitle

\section{Introduction}
\IEEEPARstart{T}{he} modern interconnected world has been constructed around a network of single-mode optical fiber. Nowadays, this foundation is beginning to crack\cite{Essiambre2010,Richardson2010,Essiambre2012}. A so-called "capacity crunch" is anticipated, whereby the single-mode  fiber systems of today will be unable to meet increasing demand - except through the expensive undertaking of adding more and more fibers.  A promising solution to this potential bandwidth crisis is spatial division multiplexing in multimode fibers\cite{Berdague1982, Winzer2012,Richardson2013,Shah2005,Stuart2000}. While standard single-mode optical fibers support just the lowest-order LP01 mode, multimode optical fibers, having a larger core, can support a multitude of transverse eigenmodes, each with different spatial shapes and propagation constants (Fig 1). Inside multimode fibers, these modes can interact through the influence of optical nonlinearity, disorder and laser gain. Although these interactions make multimode fibers much more complex than their single-mode counterparts, they also offer a world of new physics and potential applications that extends far beyond communications. 

\begin{figure*}[!t]
	\centering
	\includegraphics[width=6in]{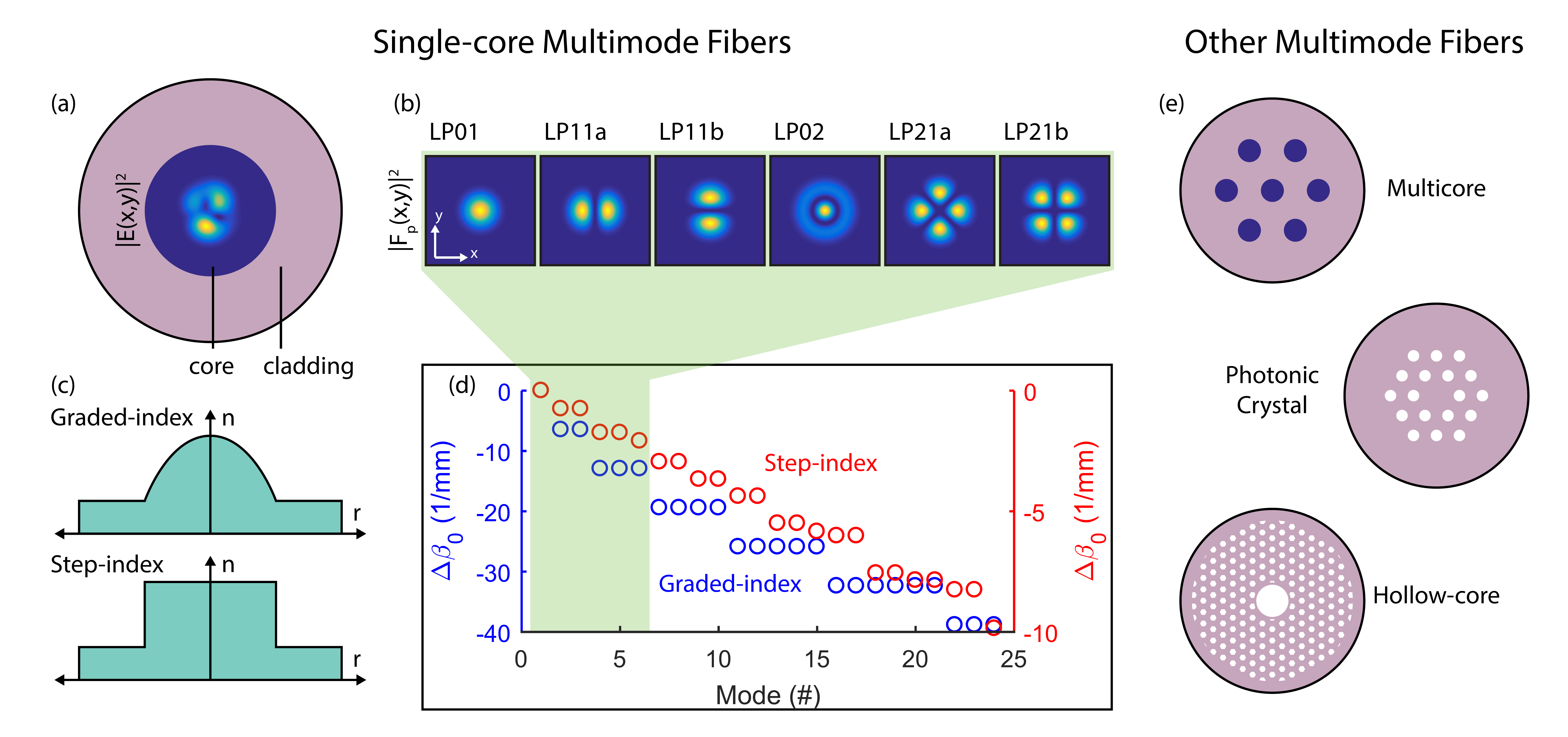}
	\caption{Modes in multimode fiber. Many studies consider fibers that support up to hundreds of modes. The linear-polarized (LP) modes are $\textbf{F}_p(x,y)=\hat{e}_{c}F_p(x,y)$, where $c = x$ or $y$ for each linear polarization. Although the LP modes are usually referred to by a pair of radial and azimuthal indices (LP01, LP11a, LP11b...), for compactness and generality throughout the rest of the text, we mainly use one index to refer to the modes, numbered starting from the fundamental mode (largest propagation constant, $\beta=2\pi n_{\text{eff}}/\lambda$). As a result of the many supported modes, multimode beams can appear quite complex, since they are superpositions of the spatial eigenmodes of the fiber. (a) shows an intensity profile $|E(x,y)|^2 = |\sum\limits_{p}^6a_pF_p(x,y)|^2$ that is a random superposition of the first 6 modes of a fiber whose intensity profiles $|F_p(x,y)|^2$ are shown in (b). (c) shows the respective index profiles of graded-and step-index multimode fibers for comparison, and (d) the corresponding propagation constants of the first 25 modes for each fiber type (with identical radius and index contrast). Step-index fibers support many more modes, all of which are close to one another in propagation constant and effective index, while graded-index fibers support discrete mode groups: modes in the same group have near-identical propagation constants, while different groups are separated by large propagation constant mismatch. (e) shows various other types of multimode fibers. More detail and information is provided in section III.B.}
	\label{fig1}
\end{figure*}

Optical fibers are the basis of several other important technologies, for which multimode fibers provide a roadmap for improvement. Today, high average power ($>$100 W) fiber lasers are a fast-growing industry. Since multimode fiber architectures spread guided light over a much larger area, they support the highest power levels and so are already widely used in these products. Femtosecond-pulse fiber lasers, which have historically only used single-mode components, are now emerging as commercial products. These devices have promise for enabling mainstream applications of ultrafast pulses, such as clinical applications of nonlinear optical microscopy and precision materials processing. For widespread adoption, these sources need to match or exceed the performance of the current standard – solid-state femtosecond lasers such as the Ti:sapphire laser – and do so at significantly lower cost. Today, fiber sources still mostly lag behind in peak power, or lack the necessary wavelength coverage. Consequently, despite the availability of reliable, low-cost instruments, many users cannot adopt the fiber platform. In this respect, multimode waveguides offer altogether new routes in generating new colors, and to scaling the peak power of these sources. 

Looking to the future, multimode waveguides can provide a new dimension in optical wave propagation, which can lead to qualitatively-new physical behaviors, and qualitatively-new kinds of applications compared to those supported by single-mode devices. For example, multimode fibers may serve as compact (even sub-mm) endoscopes, through which a laser beam can be focused, and even scanned\cite{Cizmar2011,Mosk2012,Papadopoulos2013,Ploschner2014}. Such a capability could one day enable minimally-invasive \textit{in vivo} biopsy, and biological studies deep inside scattering tissue. In addition, nonlinear optical devices based on multimode waveguides can exploit multimode and spatiotemporal processes to generate new colors or shape complex electromagnetic fields(\textit{e.g.}, \cite{Stolen1976,Hill1981,Cheng2011,Cheng2012,Pedersen2012,Mafi2012,Pourbeyram2013,Renninger2013,Ramsay2013,Demas2015,Tani2014,Wright2015,Wright2015a,Wright2015b,Demas2017,Dupiol2017,Eftekhar2017,Fang2012,Florentin2016,Guenard2017,Hellwig2014,Hellwig2016,Krupa2017,Kubat2016,Liu2016,Lopez-Galmiche2016,Krupa2016OL,Nazemosadat2016,Wright2016,Pourbeyram2015,Pourbeyram2017,SanjabiEznaveh2017,Guasoni2015,Yuan2017}).  

On the other hand, the complexity of multimode fiber optics presents major challenges. With many modes, all of which can be coupled in principle, the physics can be conceptually-difficult. As will be described, numerical calculations can be extremely helpful in isolating the roles of different physical processes, but are frequently limited by computational cost. Experiments generally require diagnostics of the spatial, temporal, and spectral content of a field, and all of the degrees of freedom may be coupled. On the other hand, multimode fibers provide a rich environment for the investigation of wave phenomena, where nonlinearity, dissipation, and disorder can be controllably included. While recent efforts provide some vindication that the complexity of multimode fibers can be understood and exploited, the science of multimode nonlinear fiber optics itself is still in its infancy.

\section{Outline and Purpose}

Our goal in this article is to provide readers the background, perspective and necessary tools for exploring multimode nonlinear optical pulse propagation. Nonlinear optics in single-mode fiber is treated exhaustively in Ref. \cite{Agrawal2007}, and basic aspects of multimode waveguides can be readily found in the literature (see for example Ref. \cite{Buck2004}). To analyze the nonlinear behavior of multimode structures, we first introduce the most common theoretical model used for this task, the generalized multimode nonlinear Schr{\"o}dinger equations (GMMNLSE, Section III). One reason we have chosen this approach is that it allows a modular dissection of all processes involved in multimode fibers\cite{Horak2012,Poletti2009}. Many limitations of this model can be reduced by adding terms that correspond to new processes, which in turn can be examined in relation to other individual effects. Hence in interpreting experiments or simulations, we can build from the ``bottom-up".

We have employed the GMMNLSE to model several experiments in our laboratory over the past few years. Due to its computational cost, it has generally necessary to restrict the calculations to a small subset of the actual modes of a fiber, and/or smaller time windows or propagation lengths than desired. This motivated the development of faster codes for multimode propagation. Here we introduce a new, massively-parallel numerical solver for the GMMNLSE. This solver is implemented for efficiency and easy editing in the MATLAB\textsuperscript{\textregistered} computing platform, and is freely available\cite{Code}. By leveraging graphical processing units (GPUs), the solver can be used for rapid modeling of multimode fiber propagation on personal computers. We demonstrate this numerical tool with several simple and concrete examples. Building on the basic concepts described by the GMMNLSE, we then describe several important effects in multimode waveguides that go beyond the GMMNLSE model as presented, such as linear mode-coupling (including disorder), and gain and loss. Finally, we highlight some important open questions for the physics and probable applications of multimode nonlinear fiber optics. 

\section{The generalized multimode nonlinear Schr{\"o}dinger equations (GMMNLSE)}

The GMMNLSE was first derived by Poletti and Horak\cite{Poletti2008}. The later, simplified version\cite{Horak2012} will be considered here, mainly for the sake of reducing the complexity of this description. While this equation has turned out to be remarkably useful, it does have many limitations that will be discussed later in the article. The GMMNLSE are a system of coupled NLSE-type equations for the electric field temporal envelope for spatial mode $p$, $A_p(z,t)$:
\begin{multline}
\partial_{z}A_p(z,t) =\\
i\delta\beta_0^{(p)}A_p - \delta\beta_1^{(p)}\partial_{t}A_p
 + \sum\limits_{m=2}^{N_d} i^{m+1}\frac{\beta^{(p)}_{m}}{m!}\partial_{t}^{m}A_p\\
+ i\frac{n_2\omega_{o}}{c}(1+\frac{i}{\omega_{o}}\partial_{t})\sum\limits_{l,m,n}^N[(1-f_R)S^K_{plmn}A_lA_mA_n^*
\\+ f_RS_{plmn}^RA_l\int_{-\infty}^t d\tau h_R(\tau)A_m(z,t-\tau)A_n^*(z,t-\tau)]
\label{eq1}
\end{multline}

We admit that this is not what would normally be referred to as a simple equation. But what we have here is not the solo performance of one mode, but instead the orchestra of a multitude of modes. This analogy is apt in that it underlines the necessity of understanding first the concepts of single-mode propagation, the single notes and rhythms, on which we have to build the chords, the counterpoint, the interacting layers of complexity within a multimode symphony. In subsequent sections we will examine each term in Eqn. 1, attempting to isolate the component concepts of multimode wave propagation, which are each expressed simply there. 

Briefly, terms 1 through 3 on the right-hand side (RHS) of Eqn. 1 are the result of approximating the dispersion operator in the mode $p$ by a Taylor series expansion about $\omega_{o}$, then transforming the terms into the time domain. The first two terms are expressed relative to the lowest-order longitudinal phase evolution ($\delta\beta_0^{(p)}$) and group velocity ($\delta\beta_1^{(p)}$) of first mode (usually the fundamental). The third term represents higher-order dispersion effects (group velocity dispersion, third-order dispersion, etc.) up to order $N_d$. $h_R$ is the Raman response of the fiber medium, $f_R \approx$ 0.18 (in fused silica) is the Raman contribution to the Kerr effect, and $n_2$ is the nonlinear index of refraction. $S_{plmn}^R$ and $S_{plmn}^K$ are the nonlinear coupling coefficients for the Raman and Kerr effect respectively\cite{Horak2012}. In keeping with our goal of simplicity, we can significantly simplify the tensors by assuming that the modes excited are in a single linear polarization, and neglecting spontaneous processes which may cause coupling into these modes. In this case, we have:

\begin{multline}
S_{plmn}^R=S_{plmn}^K=\\
\frac{\int \mathrm{d}x \mathrm{d}y [F_pF_lF_mF_n]}{[\int \mathrm{d}x \mathrm{d}y F_p^2\int \mathrm{d}x \mathrm{d}yF_l^2\int \mathrm{d}x \mathrm{d}yF_m^2\int \mathrm{d}x \mathrm{d}yF_n^2]^{1/2}}
\label{eq2}
\end{multline}
%
%

where given our last assumption, it is sufficient to express each mode in terms of a real, scalar function $F_p(x,y)$.

In many situations of interest, further simplifications can be made. Self-steepening can be neglected by taking $(1+\frac{i}{\omega_{o}}\partial_{t}) \rightarrow 1$, and stimulated Raman scattering can be ignored by letting $f_R\rightarrow 0$. We can also ignore the higher-order dispersion, considering only the group velocity dispersion. In this case, we have equations which account for many recently-observed interesting phenomena, the MMNLSE: 

\begin{multline}
\partial_{z}A_p(z,t) =
i\delta\beta_0^{(p)}A_p - \delta\beta_1^{(p)}\partial_{t}A_p
- i\frac{\beta^{(p)}_{2}}{2}\partial_{t}^{2}A_p\\
+ i\frac{n_2\omega_{o}}{c}\sum\limits_{l,m,n}^NS^K_{plmn}A_lA_mA_n^*
\label{eq3}
\end{multline}

One perspective on Eqn. 3 is worth mentioning for intuition and a complementary view of dynamics. If we undo the decomposition into modes and apply the paraxial approximation, Eqn. 3 can be written as: 

\begin{multline}
\partial_{z}E(x,y,z,t) = \frac{i}{2k_{e}(\omega_{o})}\nabla_{T}^2E - i\frac{\beta_{2}}{2}\partial_{t}^{2}E \\ +i\frac{k_{e}(\omega_{o})}{2}((n(x,y)/n_{o})^2 -1)E+ i\frac{n_2\omega_{o}}{c}|E|^2E
\label{eq4}
\end{multline}

This is the 3D NLSE for the evolution of the electric field envelope $E(z,x,y,t)$ along $z$ in the presence of a transversely-inhomogeneous refractive index $n(x,y)$. This version of the equation makes it clear that what we have is just 3D wave propagation (the LHS plus the first two terms on the RHS comprise a 3D wave equation), with perturbations from the refractive index inhomogeneity and a local nonlinear index shift. Spatial modes originate as eigensolutions to the equation including only the first and third terms (diffraction and inhomogeneous refractive index). This equation looks very similar to the Gross-Piteaevskii equation that is a well-used model for the time evolution of a trapped Bose-Einstein condensate\cite{BECbook}. It can naturally incorporate many types of effects that would be difficult to express in Eqn. \ref{eq1}. In situations where many guided modes must be considered, this equation will offer computational advantages over the MMNLSE or GMMNLSE.  When it is adequate to consider a small number of guided modes, or when the paraxial approximation is violated, it is faster to solve the coupled NLSEs. We usually find that the modal decomposition translates to higher accuracy, and to conceptually-clearer interpretation of observed dynamics. This is particularly true in practice, because the limited memory size of consumer-grade GPUs restricts the maximum variable size used to describe the propagated field. In most situations with a small number of modes, the modal basis is much more memory efficient (the number of data points needed to fully describe the field is smaller).

The following sections introduce various processes and effects that occur in a multimode fiber along with the parallel algorithm for solving the appropriate equations. The treatment is intentionally tutorial, and we emphasize that many of the presented results are well-known from prior work. We hope that this will clarify the roles of different processes for readers new to this area, and that introductory descriptions of the simplest processes will facilitate understanding of more-complex nonlinear phenomena considered later. We also hope that consideration of the simplest cases first will set the stage for description of the parallel algorithm.



\subsection{Linear propagation in multimode fibers}
\begin{figure}[!t]
\centering
\includegraphics[width=3in]{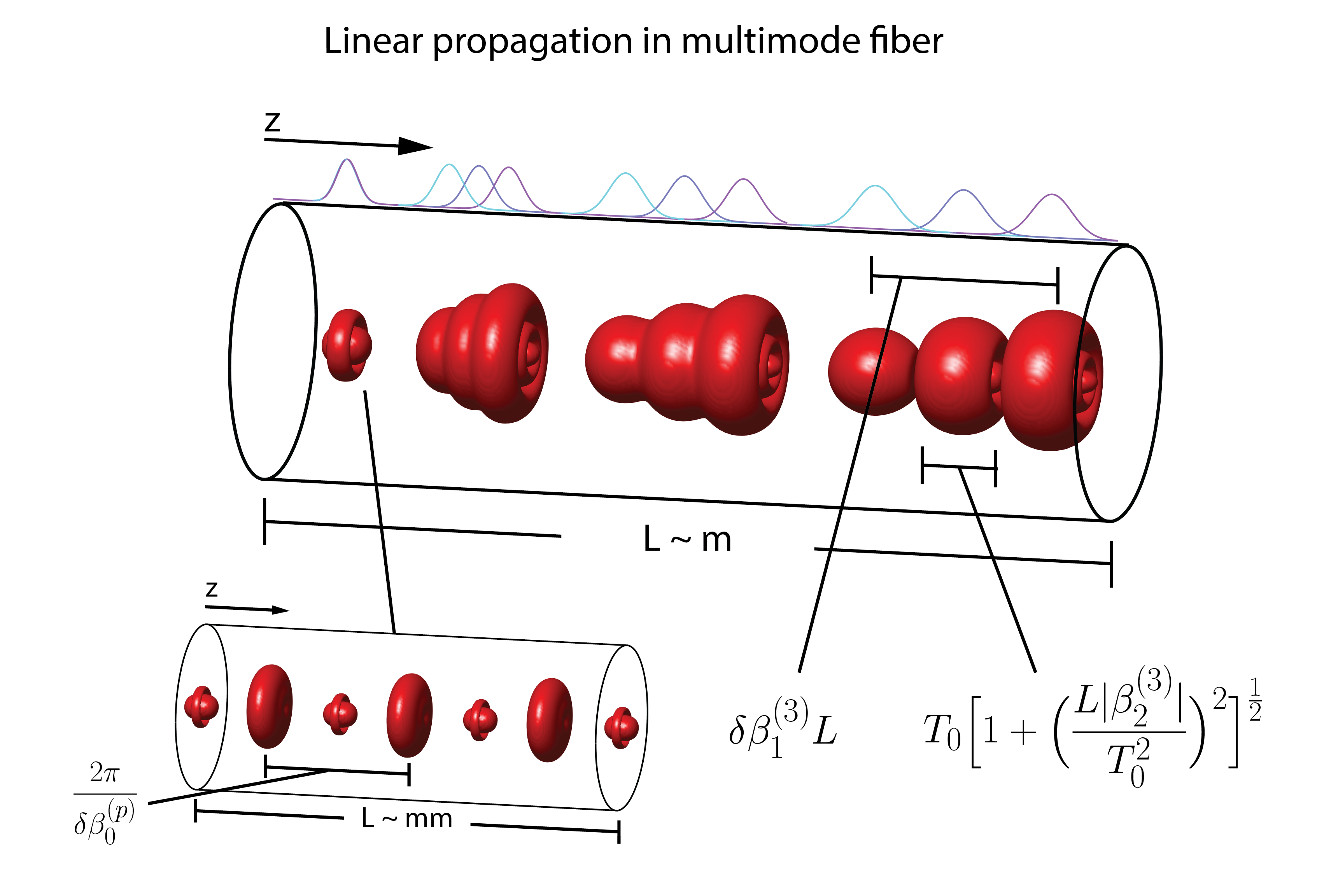}
\caption{Linear propagation effects in multimode fiber. The figure shows 3 LP0N modes, LP01 (mode 1), LP02 (2) and LP03 (3) from a parabolic GRIN fiber. The pulse in each mode broadens according due to group velocity dispersion (Eqn. 9) and the pulses in each mode move away from the pulse in the fundamental LP01 mode in time by $\delta t_p(L)=\delta\beta_1^{(p)}L$. When the pulses are overlapped, we see periodic spatial evolution of the entire multimode field, whose periodic beating depends on the first term in Eqn. \ref{eq5}. }
\label{fig2}
\end{figure}

Figure \ref{fig2} shows the main features of linear propagation in multimode fiber. These behaviors can be described by the first 3 terms in the GMMNLSE. For simplicity we let $N_d=2$, so the equations are just

\begin{multline}
\partial_{z}A_p(z,t) =i\delta\beta_0^{(p)}A_p - \delta\beta_1^{(p)}\partial_{t}A_p
- i\frac{\beta^{(p)}_{2}}{2}\partial_{t}^{2}A_p\\
\label{eq5}
\end{multline}

The eigenmodes of the multimode fiber are a set of orthonormal electromagnetic field patterns (Figure 1). Each mode has a different propagation constant, which determines the phase velocity of the electromagnetic field in that mode. There can be families of quasi-degenerate modes, such as LP11a and LP11b, which have very nearly the same propagation constant. In considering how modes interact, it can be useful to refer to the ``effective index", $n_{\text{eff}}$, that an individual mode experiences. In most fibers, the difference between the maximum and minimum refractive indices is small, and as a result the eigenmodes of the of the fiber are polarized along axes (x,y) orthogonal to the fiber axis (z). In this case, it is typical to write the modes of the fiber as ``linearly polarized" LP modes, which can be treated as quasi-scalar fields with one of two polarizations. 

As is done for single-mode propagation, the effects of chromatic dispersion are incorporated by approximating the frequency dependence of the propagation constant in each eigenmode by a Taylor series, which can then be expressed in the time domain. The pulse propagation is then examined in the reference frame of the fundamental mode (usually; another frame can be chosen if convenient), and we factor out the propagation constant - the global longitudinal phase shift - of that mode as well. Since we can only choose one reference, most of the equations in the MMNLSE have two additional terms to describe the difference in propagation constant and group velocity of each mode.  

\subsubsection{Propagation constant mismatch}

The first term in Eq. 5 represents the propagation constant mismatch, and is responsible for multimode interference or mode beating. As an example of the first term's effects, we consider multimode interference in parabolic-index fiber. In a parabolic fiber, such as is depicted in Figure 1, the propagation constants of the modes are equally spaced, with modes occupying ``mode groups" whose populations grow with decreasing propagation constant. As a result, $\delta\beta_0^{(p)} = nP$, where $P$ is the propagation constant mismatch between the mode groups, and $n$ is an integer 0,1,2,3... equal to the difference in the mode group between the two modes.   

Over short distances, the second and third terms of Eqn. 5 can be ignored. For the approximations made earlier, we can write the full electric field envelope as a composition over the modes:

\begin{equation}
 E(x,y,z,t)= \sum\limits_{p}^N\frac{F_p(x,y)}{[\int \mathrm{d}x \mathrm{d}yF_p^2]^{1/2}}A_p(z,t)
\label{eq6}
\end{equation}

Given that we are considering only the first term in \ref{eq5}, it is easy to solve for the $A_p(z,t)$ to find:

\begin{equation}
 E(x,y,z,t)=\sum\limits_{p}^N\frac{F_p(x,y)}{[\int \mathrm{d}x \mathrm{d}yF_p^2]^{1/2}}A_{po}\mathrm{e}^{i 2\pi n_pPz}
\label{eq7}
\end{equation}

The field undergoes a periodic evolution along $z$. The exact details of the periodic evolution (including the period) depend on the modes involved. In step-index or other types of fibers, this lowest-order evolution is more complex since $\delta\beta_0^{(p)}$ is not uniform. It is still generically referred to as multimode interference. An example of this rapid linear evolution, over some millimeters of fiber after a multimode excitation of 3 LP0N (radially-symmetric modes) is shown in Figure 2. We will see this lowest-order effect has a very important role when nonlinearity is considered. 

\subsubsection{Modal dispersion}
The second term in Eqn. \ref{eq5} describes modal dispersion, or modal walk-off. Each mode has a slightly different group velocity, and so a pulse which is launched into multiple modes will break up into sub-pulses whose separation grows linearly with the fiber length. If we just consider the modal dispersion term, and we assume we have some pulses in each mode with a temporal envelope $S(t)$, then the field looks like:

\begin{equation}
 E(x,y,z,t)= \sum\limits_{p}^N\frac{F_p(x,y)}{[\int \mathrm{d}x \mathrm{d}yF_p^2]^{1/2}}S(t-\delta\beta_1^{(p)}z)
\label{eq8}
\end{equation}

We therefore see the electric field envelope break up into pulses separated from the lowest-order mode pulse by $\delta t_p(z)=\delta\beta_1^{(p)}z$, whose spatiotemporal shapes are $F_p(x,y)S(t-\delta\beta_1^{(p)}z)$. \\

\subsubsection{Chromatic dispersion}
The third term in Eqn. \ref{eq5} describes chromatic dispersion.  In any particular single mode of the fiber, the effects of chromatic dispersion are the same as in propagation in a single-mode fiber: typically, short pulses broaden as they propagate. If a transform-limited Gaussian pulse of duration $T_o$  is launched into mode $p$, the pulse duration $T$ broadens as\cite{Agrawal2007,Marcuse1980}:

\begin{equation}
T(z)=T_o\sqrt{1+(z|\beta^{(p)}_2|/T_o^2)^2}
\label{eq8}
\end{equation}

In many situations that have been studied so far, the chromatic dispersion of all relevant modes is similar (i.e., $\beta^{(p)}_2$ of all modes is about the same value). However, in some cases the waveguide dispersion of particular modes (such as the highest-order modes in solid, single-core multimode fibers) can be very strong. In single-mode fibers, waveguide dispersion requires a trade-off with the effective area. As a result, dispersion engineering can make reaching high peak powers a challenge. In multimode fibers, this trade-off is eliminated, and the unusual dispersion of higher-order modes can lead to interesting and useful effects\cite{Cheng2011,Cheng2012,Pedersen2012,Rishoj2017}.

\subsection{Types of multimode fibers}
There are many types of multimode fiber. As we go on to consider nonlinear processes, it will be useful to examine how each might play out in different kinds of multimode waveguides. Scientifically, each kind of multimode waveguide provides different opportunities for nonlinear effects. For applications, each offers unique properties and advantages. First, we examine the two basic kinds of solid single-core fibers, graded-index and step-index, then multicore fibers, and then finally in more cursory detail the many other kinds of multimode waveguides to which the GMMNLSE can be applied. 

\subsubsection{Graded-index}
In a graded-index (GRIN) fiber, the refractive index smoothly changes through the core, reaching a maximum at the center (Fig. 1(c)). Usually the profile is optimized to be close to a parabola. As shown in Fig. 1(d), in a GRIN fiber the modes cluster into nearly degenerate groups. Modes that belong to the same group have minimal modal dispersion between one another. As a result, the modal dispersion in graded-index fibers is the minimum possible with a single-core design. The magnitude of the modal walk-off parameters within a given mode group in GRIN fiber can be about 10-100 times smaller than those in step-index fibers of similar specifications (radius, index contrast, etc.), and the sign can even be controlled by slightly modifying the design from a perfect parabola\cite{Li2012}. In practice, this means that one can observe strong nonlinear intermodal interactions with very short ($\sim$100 fs) pulses. As seen above, the regularity of the mode spacing means there is a strong periodic component to the longitudinal evolution of the field. This has been the source of several interesting observations, such as resonant radiation from oscillating multimode solitons\cite{Wright2015a}, and the geometric parametric instability\cite{Longhi2004,Krupa2016,Lopez-Galmiche2016,Krupa2016OL,Tegin2017}. While intragroup modal dispersions are very low, the effective index spacing between adjacent mode groups in GRIN fiber is actually quite large - larger, in fact, than the spacing between adjacent modes in a similar step-index fiber.  Linear mode coupling (such as from disorder in the fiber, see Section F1) decreases rapidly with the difference in propagation constants, so the fundamental mode of a graded-index fiber is the most stable mode of {\it{any}} multimode fiber (although the situation becomes much more interesting when nonlinearity is considered\cite{Wright2016}).

\subsubsection{Step-index}
In a step-index fiber (Fig. 1(c)), there is not quite as much regularity in the mode structure as in GRIN fiber, and the modal dispersion is usually much stronger. The first feature has proven useful, for example, in exciting high-order modes in these fibers (primarily the LP0N family). The radial symmetry and the large difference in propagation constant from nearby modes mean that these modes are weakly affected by disorder and can propagate over long distances\cite{Ramachandran2006}. Compared to GRIN fibers, there are fewer degeneracies. A step index fiber supports more modes than a GRIN fiber with similar specifications. For example, the GRIN and step-index fibers shown in Fig. 1(d) have equal core radius and index contrast, but support 170 and 330 spatial modes, respectively.  The effective area of modes in step-index fiber tend to be larger than in GRIN fiber. Nevertheless, unlike in GRIN fiber the effective area of high-order LP0N modes actually decreases with mode order. 

The strong modal dispersion in step-index fibers poses a challenge for observing interactions between short pulses in different modes, especially with anomalous dispersion in any of the modes. To date, multimode dynamics in step-index fibers have been primarily limited to broadband intermodal four-wave mixing processes (e.g., \cite{Stolen1974,Stolen1976,Pourbeyram2015,Pourbeyram2017,Demas2015,Demas2017}). However, using longer pulses and particularly in the normal dispersion regime, we expect that step-index fibers should support a wider range of behaviors that involve other processes. An example is considered in Section E3.

\subsubsection{Multicore fibers}
Multicore fibers are multimode fibers formed by an array of separate single-core fibers (the cores may be single-mode, multimode, graded-index, or step-index, etc.). An example with 7 cores is depicted in Fig. 1e. In the context of nonlinear multimode dynamics, multicore fibers are another class of fiber that is under-explored. When the cores are strongly coupled (\textit{i.e.}, a few diameters apart), the modes of a multicore fiber are the so-called ``supermodes". These can be interpreted as hybridizations of the individual modes of the separate cores. The supermodes have qualitatively different dispersive properties than the modes of the individual cores. The properties of multicore fibers can be engineered much more easily than single-core multimode fibers. The modal dispersion can be similar to GRIN fibers, and waveguide dispersion can be strong even for the lowest-order modes. The effective area of the supermodes can be large, and the overlap between the different supermodes can be adjusted. When the cores are relatively weakly or moderately coupled, an N-core multicore fiber is sensitive to disorder, since the supermodes of the fiber are very nearly degenerate\cite{Arik2013,Mumtaz2013}. Finally, multicore waveguides are attractive for their potential ease of integration with multiple single-mode fibers.

\subsubsection{Other multimode waveguides}
Adding spatial modes is a technique that can be applied to virtually any optical guided-wave system to achieve new and/or better behaviors. For example, solid-core and hollow-core photonic crystal fibers (Fig. 1e) can guide multiple modes. In hollow-core photonic crystal fiber, interactions between spatial modes have been used for phase-matching many unique processes\cite{Russell2014,Tani2014} – however to date only fibers with a small number of modes have been considered. Waveguides written into planar or bulk media (fused silica, or SiN, for example)\cite{Davis1996,Khurmi2016} may also support a few or many spatial modes.  Many microresonators (and ``macro" resonators like Herriot cells\cite{Schulte2016,Hanna2017}) support multiple spatial modes, and the opportunities afford by multimode processes have just recently begun to be explored in these systems\cite{Xue2015,Liu2014,Yang2016,Guo2017,DAguanno2016}. Last, multimode ``rod fibers" fabricated with very large claddings, or with very large dimensions offer an opportunity for studying highly multimode dynamics. By making the waveguide inflexible, these rods would sacrifice some practical benefits of fiber. However, for multimode operation this inflexibility may be very important for realizing environmental stability.

\subsection{Nonlinear propagation in multimode fibers}
Nonlinearity underlies many surprises, opportunities, and problems in multimode propagation.  Fundamentally, nonlinearity introduces coupling between the different modes, and is incorporated in the last term(s) of the GMMNLSE and MMNLSE.

\subsubsection{Self phase modulation}
Self-phase modulation (SPM) corresponds to $S_{pppp}^K$ and terms of the form $i|A_p|^2A_p$. The fact that $S_{pppp}^K$ is distinct for each mode can lead to some interesting outcomes. Modes have different effective areas ($A_{\text{eff}}=1/S_{pppp}^K$). The effective area can be very large, so for applications such as high-power lasers, this means that higher power can be guided before nonlinear distortions become an issue. The differential rate of nonlinear phase accumulation can also be important for intermodal processes, since it leads to a decorrelation of the modal phases.  An example is the observation of irreversibility without dissipation that occurs in Kerr beam cleanup, where unequal self-phase modulation arrests the periodic backconversion of energy through four-wave mixing\cite{Krupa2017}. 

\subsubsection{Cross phase modulation}
Terms with $S_{pnpn}^K$ or $S_{ppnn}^K$, where $n \neq p$, are of the form $ i|A_n|^2A_p$. These are cross-phase modulations (XPM). The role of these terms in the formation of multimode solitons is examined in the example in Section E2. XPM can lead to asymmetric spectral broadening when two pulses are traveling at different speeds. SPM and XPM are pure phase modulations, and cannot cause energy exchange between modes. 

\subsubsection{Four wave mixing} 
All other nonlinear coupling terms can be described as ``four-wave mixing" (FWM), which we will define as ``terms that can cause transfer of energy". Examining the nonlinear terms in the GMMNLSE, we see that all terms are technically ``four-wave mixing" in that they involve a mixture of four, possibly different, waves. But within our definition, four-wave mixing terms are ones that can take on complex values. We can understand FWM as the scattering between four waves due to the nonlinear index modulations created by their interference. Significant energy exchange in a FWM process occurs only when energy and momentum are conserved (the latter usually being referred to as phase matching, and expressed in terms of the phase velocity of the interacting waves). In MMF, each of the four interacting modes can have a different frequency and be in a different spatial mode, so the conditions for FWM processes (except third-harmonic generation) are that:


\begin{multline}
\omega_1+\omega_2-\omega_3-\omega_4 = 0;\\
\Delta\beta=\beta_3(\omega_3)+\beta_4(\omega_4) -\beta_1(\omega_1)-\beta_2(\omega_2)+\Delta \beta_{NL}= 0;
\end{multline}

Here, the final term in $\Delta\beta$ is a nonlinear contribution to the propagation constant. In most cases, this has a small influence on broadband multimode processes

Practically, the phase-matching requirements show that there are many degrees of freedom for controlling a FWM process in a highly-multimode fiber. We may be interested in controlling the generated frequency, the generated mode, the bandwidth of the four-wave mixing process, or some features of the generated photon's correlations/entanglement. By adjusting the spatial mode, and frequency of two or three of the waves involved, and/or by engineering the dispersion of the modes involved by design of the fiber, one has a significant degree of control over the modes or frequencies of the remaining waves. One can also control the rate of change of $\Delta\beta$ near 0, so as to affect the bandwidth, or range of frequencies, generated by the FWM process. 

\subsubsection{Self-steepening}
The self-steepening terms are proportional to $-\frac{n_2}{c}\partial_t$. As in single-mode fiber, self-steepening in a multimode settings is expected to lead to the formation of an increasingly steep trailing edge (i.e., the side of the pulse where the t-variable in the GMMNLSE is highest) and the corresponding broadening of the blue side of the spectrum. The self-steepening terms may also be responsible for intermode energy transfer, since $\partial_tA_lA_mA_n^*$ may be complex. To date, self-steepening has not been responsible for any major features of propagation in multimode fiber. However, it regularly has a noticeable effect and it is often worth repeating simulations with and without this term to examine how it affects the results. In single-mode fiber, one would estimate self-steepening to be important when the fiber considered is comparable to the shock distance, $\approx 0.39z_{NL}\omega_0T_0$, where $z_{NL}=A_{\text{eff}}/(n_2\omega_0P_0)$ is the nonlinear length, and where $P_0$ is the peak power, and $T_0$ is the pulse duration. This will usually be a good guideline in multimode fiber too. However, like many such characteristic lengths, it is only a loose guideline for how the solution to the GMMNLSE behaves. 
\subsubsection{Raman scattering}
Raman scattering is a dissipative process where a phonon from the fiber medium interacts with the electric field, causing a spectral redshift that is either continuous (in the case of solitons and multimode solitons\cite{Wright2015,Zhu2016}) or discrete (more typical in the normal dispersion regime\cite{Pourbeyram2013,Wright2016}, but also observed for group-velocity matched modes with anomalous dispersion\cite{Yang2016,Rishoj2017}). Raman scattering underlies multimode spectral incoherent solitons\cite{Fusaro2016}. It can cause energy transfer between modes, which occurs for example in Raman beam cleanup\cite{Terry2007}. Within the GMMNLSE, stimulated Raman scattering is modeled using a phenomenological medium response function, $h_R$. 

\subsection{Parallel algorithm for solving the GMMNLSE}
The GMMNLSE can be solved using the same numerical schemes used to solve the NLSE and GNLSE. Aspects and implementation of the numerical solution to the GMMNLSE have been considered by Poletti and Horak \cite{Poletti2008}, and later Khakimov \textit{et al.}\cite{Khakimov2013}. The most-common approach is based on a split-step (SS) algorithm, where the nonlinear terms are integrated in the time-domain and the linear dispersive terms are evaluated as multiplications in the spectral domain. For the GMMNLSE, this approach works well enough, but the nonlinear step quickly becomes the limit as the number of modes increases. For each mode $p$, propagating one step requires calculating three nested sums, each over all modes. The overall computational complexity associated with this is therefore $O(P^4)$ for $P$ modes, so the computational complexity grows extremely quickly for more than a few modes. For typical resolutions, we find that this usually means the GMMNLSE is the most efficient model when the number of modes is smaller than about 10-30, beyond which the full-field techniques such as the (3+1)-D GNLSE or the unidirectional pulse propagation equation\cite{Andreasen2012,Eftekhar2017} become increasingly advantageous.

Here, we solve the GMMNLSE using a Massively Parallel Algorithm (MPA)\cite{Lushnikov2001,Lushnikov2002,KorotkevichLushnikov2011}. This is still subject to the poor scaling of the nonlinear term, but uses a parallel method to speed up the calculations. In a typical split-step solution to the GMMNLSE, the equation is separated into the linear and nonlinear terms, and integrated across each small step size in $z$ over which these effects are assumed to be approximately independent. The same is done over the next step in $z$, in essence ``beam propagation" along $z$. In order to leverage the increasingly parallel capabilities of modern computers, however, the MPA considers many subsequent small steps, which are computed in parallel and then iterates the until the error is below a given tolerance. For each step, the linear term is solved exactly, while the nonlinear term is numerically integrated, much like with the split-step method. As long as the number of iterations is smaller than the extent of parallelization, which can always be the case, this algorithm can be faster than a more traditional serial split-step algorithm.

Before continuing to explain the MPA, it is worth pointing out that the MPA provides a means of parallelizing many similar equations/numerical schemes, including the (3+1)-D GNLSE, and the methods recently developed by L\ae gsgaard\cite{Lægsgaard2017,Kolesik2004} and Conforti et al.\cite{Conforti2017}. These latter schemes are not discussed in more detail here due to their recency. Briefly, the the former approach appears to provide a significant benefit for highly-multimode simulations, both in terms of accuracy and speed, and the latter should provide very large speed-ups, albeit with good accuracy only for quite restricted conditions. Of course, the GMMNLSE reduces for one mode to the GNLSE. Since for one mode propagation is not limited by a short beat length, the MPA is less useful. 

The numerical codes released in conjunction with this article include an implementation of the MPA for the GMMNLSE as described in earlier sections\cite{Code}, as well as the split-step implemented with and without GPU functionality, and multiple tools for multimode fiber calculations, such as a mode solver from the MATLAB\textsuperscript{\textregistered} file exchange\cite{Fallahkhair2008}. In a future work, we hope to incorporate a broader set of codes, which will be discussed later in this article.

Here we provide an overview of the MPA. A more systematic description of the MPA GMMNLSE algorithm is included as part of the documentation for the numerical code package \cite{Code}, as well as within the commenting of the code itself. Starting with the GMMNLSE, we group the dispersion term (including mode-beating, modal and chromatic dispersions) into the linear operator $D(t)=i\delta\beta_0^{(p)} - \delta\beta_1^{(p)}\frac{\partial}{\partial t} + i\sum_{n \ge 2}{\frac{\beta_n^{(p)}}{n!} \left ( i \frac{\partial}{\partial t} \right )^n}$ and using the notation $F[D(t)A_p(t, z)] = D_\omega A_p(z, \omega)$, where $F$ represents the Fourier Transform, MPA defines the change-of-variables

\begin{gather} \label{cov}
A_p(z, \omega) = \psi_p(z, \omega) \textrm{exp}[D_\omega (z-z_0)]
\end{gather}

Here $\psi_p(\omega,z)$ would be independent of $z$ for purely linear propagation. Adding nonlinearity makes $\psi_p(\omega,z)$ a slow function of $z$ since the nonlinearity is weak. Applying the change of variables, transforming to the Fourier domain, and integrating converts the GMMNLSE, Eqn. 1, into

\begin{dmath} \label{MPA_eq}
	\psi_p(z,\omega) = \psi_p(z_0,\omega) +\\ i \frac{n_2 \omega_0}{c} \left(1+\frac{\omega}{\omega_0}\right)
	\int_{z_0}^z \sum_{l,m,n} F \left \{ (1-f_R)S_{plmn}^K A_l(z^\prime, t) A_m( z^\prime, t) A_n^*(z^\prime, t) + f_R S_{plmn}^R A_l(z^\prime, t)\int ^t_{-\infty}h_R(\tau) A_m (z^\prime,t-\tau)A_n^*(z^\prime, t-\tau)d\tau  \right \}\textrm{e}^{-D_\omega (z^\prime-z_0)}dz^\prime
\end{dmath}

It should be noted that this is still an exact form of the GMMNLSE, no discretization has been applied yet. From this form, however, one can see that the solution could be computed efficiently in parallel if the space along $z$ were broken up into a number of discrete steps and at each point $z^\prime$ the integrand was computed in parallel. This is beneficial because by a large margin the most computationally expensive calculation is the $O(P^4)$ sum, which exists in the integrand. Because of the nonlinear nature of the integrand, in general such a parallelization can not be done; however, by inspecting the length scales of the problem we can formulate a solution.

We begin by defining two step sizes along $z$. A large step size, $L$ is broken up into $M$ small steps $\Delta z$ in order to compute the integrand in parallel. The two longitudinal step sizes, the small step size $\Delta z$ and the longer step size $L= M\Delta z$, are determined by the two important length scales in the problem, the nonlinear length, $z_{NL}$, and the intermode beat lengths, $z_{IM}=1/\delta\beta^{(p)}_o$. The nonlinear length depends on the intensity, but is often no less than 50-100 $\mu$m and frequently is orders of magnitude larger. While we typically estimate $z_{NL}=A_{\text{eff}}/(n_2\omega_0P_0)$ as in single-mode fiber, it should be noted that, due to mode interference (e.g., the inset in Fig. 2), the length scale for nonlinear effects in MMF often differs significantly from the single-mode estimate. The beat lengths between the modes depends on the fiber itself, however, and can be as small as a few $\mu$m. Given these two lengths, therefore, we can choose the large step size $L$ such that $L \ll z_{NL}$ and the small step size $\Delta z$ such that $\Delta z \ll z_{IM}$. Of course, these must be related by $M\Delta z=L$ (i.e. M small steps fit into one large step). The $z$ grid can then be denoted as the set of equally spaced points $z_0$, $z_1$, ..., $z_M$, where $z_M=z_0+L$.

Each parallel step is computed as follows. Over a distance of $L$, $\psi_p(z, \omega)$, which represents the result of nonlinear phase accumulation, is almost constant. To a first approximation, therefore, MPA sets $\psi^{(n=1)}_p(z_j, \omega)=\psi_p(z_0, \omega)$, where $n$ is the iteration number, for $j = 1...M$, which also establishes $A^{n=1}_p(z_j, \omega)$ through Eqn. \ref{cov}. Next each integrand is calculated in parallel, and then summed to get a more accurate approximation of $\psi_p(z_j, \omega)$. This process is then repeated in an iterative fashion. Each iteration, the algorithm recalculates $\psi^{(n)}_p(z_j, \omega)$ by computing each integrand in parallel   and then summing over $z'$ using the by trapezoid rule. $\psi^{(n)}_p(z_j, \omega)$ is converted to $A^{(n)}_p(z_j, \omega)$ through Eqn. \ref{cov}, and then the process repeats, each time giving $A^{(n)}_p(z_j, \omega)$ with a higher and higher accuracy. Once the accuracy reaches a given threshold, the process is considered converged and $A^{(n_{tot})}_p(z_M, \omega)$ is taken as the final value at the end of the large step. This finishes one step from $z_0$ to $z_0 + L$, after which the next large step $L$ can be taken using $A^{(n_{tot})}_p(z_M, \omega)$ as the new $A_p(z_0, \omega)$. The relative error from such iterations $\sim(L/z_{nl})^{n+1}$ as found in Refs. \cite{Lushnikov2002,KorotkevichLushnikov2011}.    

In practice, even if $L \approx \frac{1}{10} z_{NL}$ this method converges to a high degree of accuracy in $n_{tot}=2-3$ iterations at most. Therefore, it is typically the case for simulations in MM fiber that the MPA provides a significant parallel speedup. It should be noted that such a speedup depends not only on $n_{tot}$ being much smaller than $M$, but also heavily on the effectiveness of the implementation of parallelization. If $M=10$ but computing the integrand for 10 points ``in parallel" computes each at 1/10th the speed that it would take to compute a single integrand in serial, this algorithm will not be effective and in fact it will be slower than a split-step method due to the added overhead. As a result, a high degree of parallelization is important, which motivates our use of GPUs.

Figure 3 shows the time comparison between the different algorithms, all implemented in MATLAB, for a simulation of 10 modes of a graded-index fiber, under conditions for which the ratio $z_{nl}/\Delta z \approx$ 200, which is typical of our usage.  For the MPA, we vary the $M$ parameter. For our system (a high-end PC, with the top modern commercial GPU), simply using the GPU creates a significant speed-up. This is not surprising given the numerous summations required. Above this, MPA provides another speed-up, approaching an additional order-of-magnitude. As $M$ increases, $M\Delta z$ approaches $z_{nl}$ and increasingly higher values of $n_{tot}$ are required for convergence over each large step. Hence, the value of $M$ for which maximum speed-up occurs depends on the ratio of $z_{nl}$ to the required small step size $\Delta z$, as well the overhead of the MPA relative to the split-step. The code used to generate Figure 3 is included with the numerical code package. We verified the accuracy of the MPA code by comparison to analytic expressions, such as from Ref. \cite{Agrawal2007}, to older codes in our group for MM and single-mode propagation, and to predictions of the 3D NLSE, Eqn. 4. 

\begin{figure}[!t]
	\centering
	\includegraphics[width=3in]{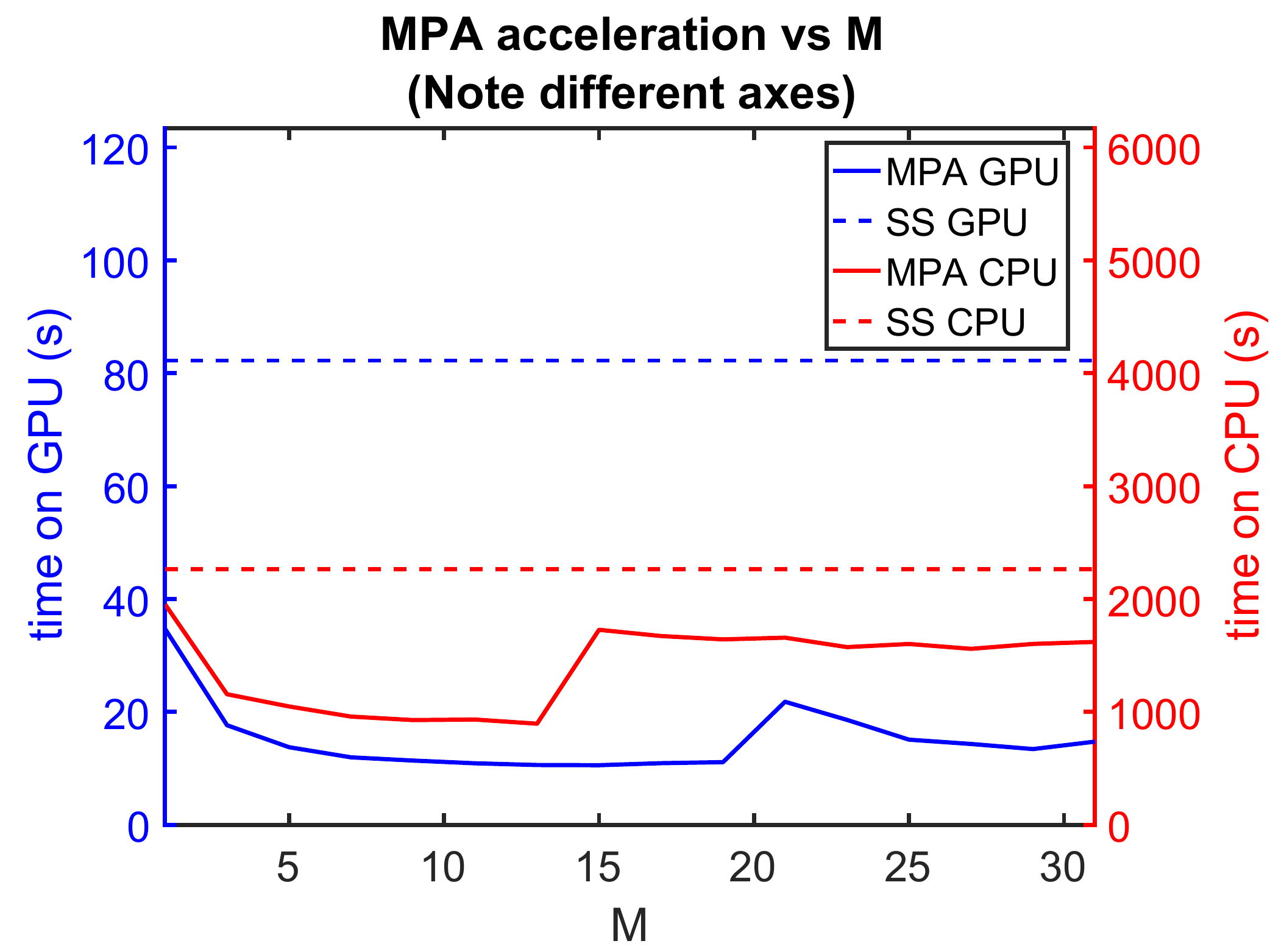}
	\caption{Time comparison for simulations with the GMMNLSE solved using split-step method (SS) and the MPA with the same numerical accuracy.  For each case, we consider varying parallelism in the MPA (parameter $M$). A significant speed-up is also observed from utilizing GPU functionality within MATLAB.}
	\label{fig4}
\end{figure}

\subsection{Examples calculated with the GMMNLSE}
In the numerical code package available online, we have included three examples that walk through representative studies of pulse propagation with the GMMNLSE. We summarize the findings of these studies here, but their main purposes are to illustrate 1) the use of the various codes (including calculation of the modes\cite{Fallahkhair2008}, dispersion and coupling tensors), 2) the use of the modularity of the GMMNLSE to understand complex nonlinear pulse propagation, 3) to emphasize the equation's limits, and where to be cautious about its accuracy, and 4) to provide some some guidelines for avoiding and identifying numerical artifacts.
\subsubsection{Example 1: linear propagation in a multimode fiber}
In this example, a short pulse is launched into all the modes of the fiber, with very small peak power. The results are similar to those shown in Fig. 2. 
\subsubsection{Example 2: Multimode soliton formation}

\begin{figure}[!t]
	\centering
	\includegraphics[width=3in]{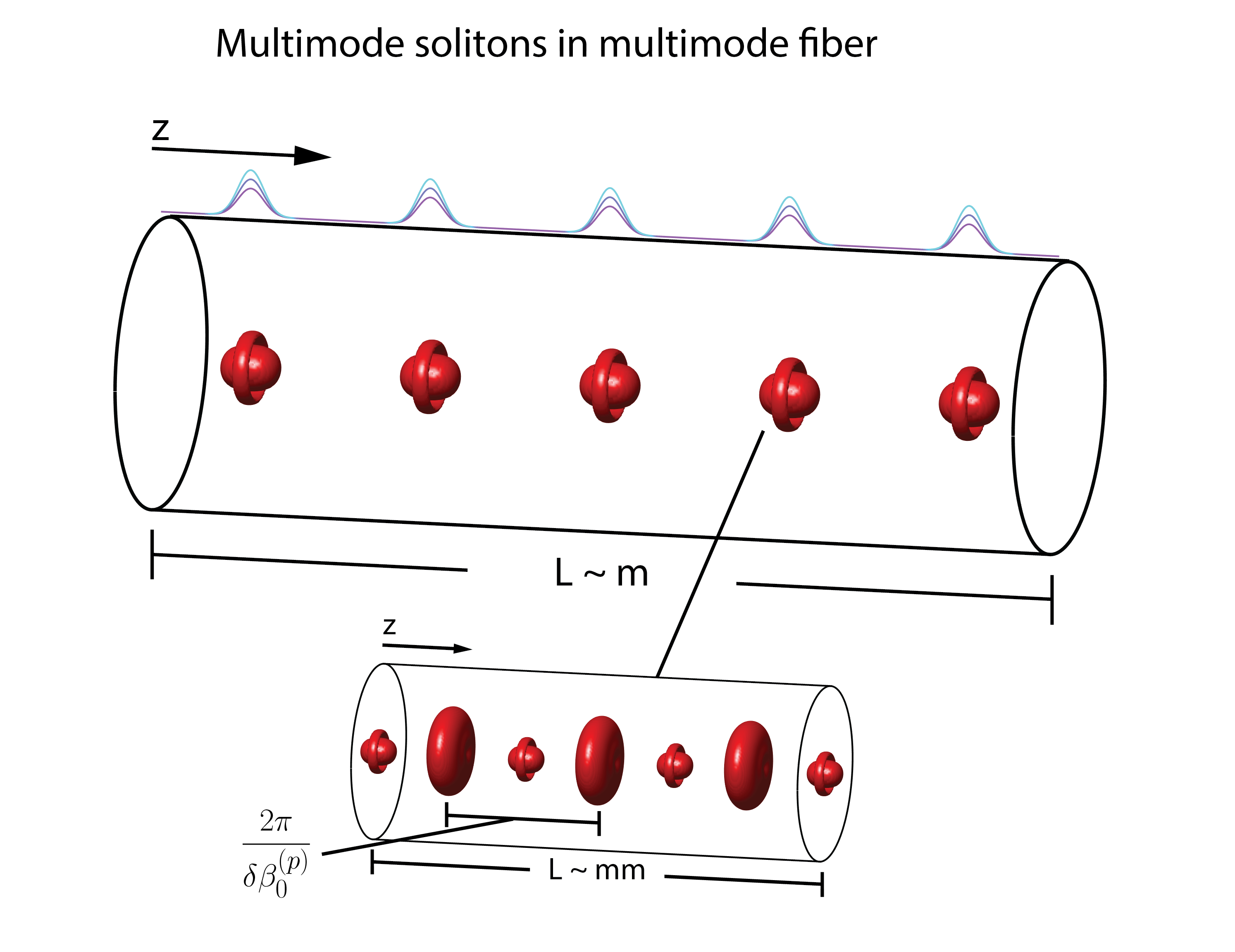}
	\caption{Soliton propagation in multimode fiber. The figure shows 3 LP0N modes, LP01 (mode 1), LP02 (2) and LP03 (3) from a parabolic GRIN fiber. Multiple spatial modes lock together in a single pulse, which evolves over short distances due to the different modal propagation constants. The formation of MM solitons is considered in Example 2.}
	\label{fig3}
\end{figure}

In this example, we consider propagation in a GRIN fiber modeled after standard telecommunication-grade GRIN multimode fiber, and look at the formation and propagation of multimode solitons\cite{Hasegawa1980,Crosignani1981,Yu1995,Raghavan2000,Renninger2013,Wright2015,Wright2015a,Zhu2016,Buch2015,Buch2016,Mecozzi2012}. Fig. 4 illustrates the basic concepts underlying multimode solitons. In contrast to the linear propagation shown in Fig. 2, with nonlinearity we may observe a multimode pulse that does not broaden in time, nor break apart into multiple pulses in each constitutent mode family. Anomalous chromatic dispersion is balanced by self-phase and cross-phase modulations.

Since the modes have different group velocities, we expect that cross-phase modulations cause asymmetric spectral broadening of pulses in different modes. This process can shift the average frequency of the pulse in each mode, and therefore cause pulses in the different modes to have a common group velocity, where the changes in chromatic dispersion compensate the mismatch in modal dispersion. Intuitively, this process can be thought of as a mutual trapping of each pulse by the others and itself.

Hence, as illustrations of use of the codes for numerical investigations, in this example we roughly aim to answer the questions ``What is the role of the initial pulse duration on the excitation of highly multimode solitons?", and ``Is cross-phase modulation the only important process for the formation of the multimode soliton?"

We consider the first 6 modes of the fiber, plus 2 additional LP0N modes. In our first simulation, we launch 6 nJ, equally distributed among those modes, and monitor propagation through 15 m. With a 50-fs initial pulse duration, a pristine multimode soliton emerges and experiences a soliton self-frequency shift. Similar to experimental work\cite{Wright2015,Zhu2016}, we filter out the shifted soliton in order to examine it in isolation (Fig. 5a-b). With Raman and four-wave-mixing terms included, we see that energy shifts from higher-order to low-order modes, also in agreement with experimental findings in this kind of fiber. As expected due to XPM, the spectrum in each mode shifts slightly to help equalize the group velocity across the modes. When we neglect self-steepening and Raman, and include only the Kerr SPM and XPM terms, a MM soliton also forms for the same initial condition (Fig. 5c-d), which illustrates that XPM alone is sufficient for MM solitons to form. With all terms included, a 1-ps initial pulse breaks up into multiple pulses, and at 15 m propagation no clear MM soliton has formed (Fig. 5e-f). However, at around 3 ps in Fig. 5e, we see evidence of a MM soliton comprising a subset of the launched modes (see inset Fig. 5f). If the reader cares to extend the propagation to well beyond 15 m, they will observe formation of several MM solitons, each with a unique (and perhaps surprising) composition. In summary, we can answer our earlier questions as ``a shorter pulse can produce a more-multimode soliton," and ``qualitatively yes, but quantitatively no."

\begin{figure}[!t]
	\centering
	\includegraphics[width=3.5in]{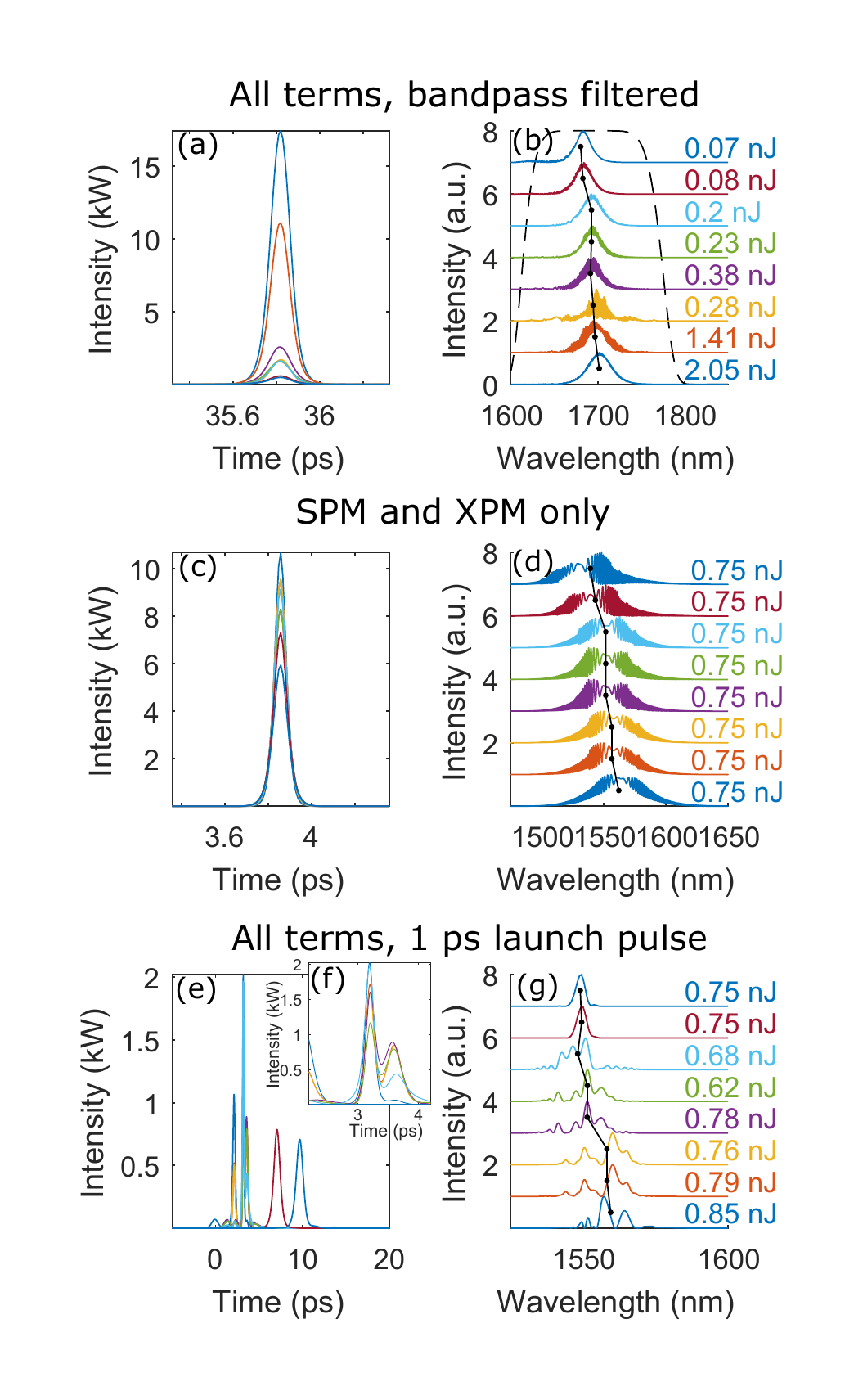}
	\caption{Examination of multimode soliton formation and propagation. (a-b) show the propagated field after 15 m and bandpass spectral filtering to isolate the Raman-shifted soliton. The pulse is initially 50 fs long, with 6 nJ equally distributed into the 8 modes considered. (c-d) show the result of propagation for the same initial condition, but with only the Kerr SPM and XPM terms included. (e-g) show the same simulation as (a-b), except with a 1 ps duration initial pulse. The inset, f, shows a zoom-in of e to emphasize a forming multimode soliton. The energy values plotted in b, d, and g refer to the total energy within the respective spatial mode plotted (starting with mode 1, the fundamental mode, at the bottom).}
	\label{fig5}
\end{figure}

\subsubsection{Example 3: Generation of 1300-nm pulse through self-phase modulation}
For biomedical nonlinear optical microscopy, excitation at 1300-nm is highly desired\cite{Horton2013}. Considering absorption and scattering together, this window corresponds to an attenuation minimum in typical tissue. Since well-developed fiber lasers emit around 1030 nm, 1550 nm, or 1900 nm, nonlinear frequency conversion has been explored to reach 1300 nm. One simple approach has been to apply self-phase modulation to transform-limited pulses in short fibers\cite{Liu2016b,Liu2017b}. If the propagation is dominated by SPM, we expect well-defined spectral sidelobes. Since the outermost sidelobes correspond to inflections in the phase of the pulse, when filtered by a bandpass filter, they are nearly transform-limited pulses\cite{Liu2016b,Liu2017b}. In order for this filtered sidelobe to have a high pulse energy, we'd like to be able launch a very high energy pulse, and in a reasonable length of fiber (several centimeters at least) accumulate the spectral broadening. In this example, we are concerned with the question ``Can spectral broadening of a pulse launched into the fundamental mode of a graded-index or step-index fiber provide a Gaussian pulse at 1300 nm?"

Figure 6 shows the results of this investigation. Briefly, we find that there is strong energy transfer due to the high peak powers. In GRIN fiber, energy is relatively stable within the fundamental mode, while in the step-index fiber (which we design to have a similar fundamental mode area), significant energy is transferred into the LP02 mode (\textit{i.e}, mode 6 in Fig. 6). Both these observations can be understood simply in terms of the effects of self-focusing (although the peak power here is below the critical power for collapse, non-catastrophic self-focusing nonetheless occurs). We see that energy transfer in the GRIN fiber occurs primarily on the trailing edge of the pulse, while in the step-index fiber, substantial energy is also transferred at the peak. In both cases, however, an energetic pulse can be obtained at 1300-nm. The pulses are slightly chirped. Given the very high peak power here, readers should question if such results could be obtained experimentally without damaging the fiber. Nonetheless (and in part \textit{because} of this question), the problem is a good example of exploration using the GMMNLSE. Within the context of these important uncertainties, we can answer our earlier question with ``yes". 

\begin{figure}[!t]
	\centering
	\includegraphics[width=3.5in]{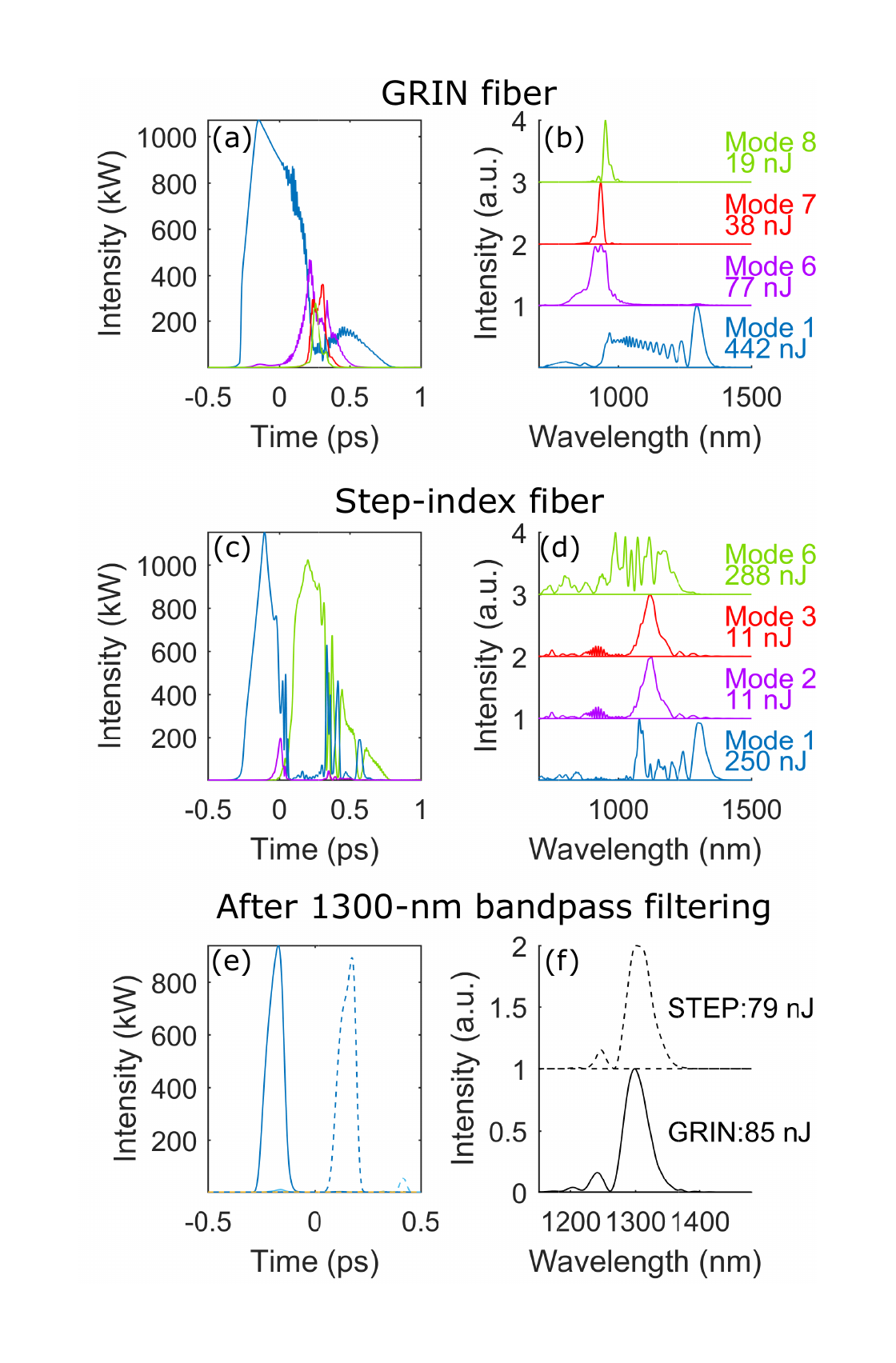}
	\caption{Generation of $\sim$ MW pulses at 1300-nm by self-phase modulation in the fundamental mode of GRIN and step-index multimode fibers. (a-b) show the result of propagation after 3.6 cm for a 600-nJ, 200 fs pulse launched into the fundamental mode of a GRIN fiber. (c-d) show the same for 2.8 cm step-index fiber. For clarity, in (a-d), we are only showing modes which have more than 1$\%$ of the total energy. (e-f) show the result after bandpass filtering the above fields at 1300-nm. Nearly-transform-limited pulses in the fundamental mode are obtained.}
	\label{fig6}
\end{figure}

\subsection{Other effects that can be treated within the GMMNLSE}
\subsubsection{Linear mode coupling including disorder}
Any small deviation from the ideal waveguide (for which the modes used in the MMNLSE were calculated) can be modeled as causing coupling between the ideal guided modes of the unperturbed fiber, \textit{i.e.} perturbative coupled mode theory. The GMMNLSE can then be supplemented by an additional term on the right hand side\cite{Mumtaz2013,Xiao2014}, $C= i\sum_n^NQ_{np}A_n$, where the coupling coefficients are given by:

\begin{dmath} \label{MPA_eq}
Q_{np}(z)=\\\frac{k_o}{2n_{\text{eff}}}\frac{\int \mathrm{d}x \mathrm{d}y [n^2(x,y,z)-n_p^2(x,y,z)]F_n(x,y)F^{*}_p(x,y)}{[\int \mathrm{d}x \mathrm{d}yF_n^2\int \mathrm{d}x \mathrm{d}yF_p^2]^{1/2}}
\end{dmath}

In this equation, $n_p(x,y,z)$ is the perturbed index profile and $n(x,y,z)$ is the ideal profile. If we consider just the longitudinal component of this, taking $[n^2(x,y,z)-n_p^2(x,y,z)]=\epsilon(x,y)f(z)$, then the Fourier transform $\tilde{f}(k)$ of the longitudinal component tells what longitudinal “momentum” the medium perturbations are providing to coupling processes. In other words, $\tilde{f}(k)$ has a large component at some $k=\delta\beta$, then two spatial modes whose propagation constant separation is $\delta\beta$ will be phase matched through the perturbation for efficient coupling. For example, in the case of a long-period Bragg grating, a non-zero virtual momentum corresponding to the grating period can allow coupling between modes whose beat lengths are near that period.

In the case of disorder, $f(z)$ is reasonably assumed to be a Gaussian stochastic function. This disorder would typically not have any particular periodicity, so  $\tilde{f}(k)$ is large only near $k=0$ and therefore we expect that only modes with similar propagation constants are coupled together by disorder. In single-core fibers these are the quasi-degenerate mode groups. The disorder is also usually described by its correlation length, which refers to the distance through which index perturbations, or equivalently linearly propagating fields in the coupled modes, become uncorrelated. Most estimates for this length in typical glass fibers are in the 10-100 m range \cite{Agrawal2007}. 
	
When the correlation length is shorter than the fiber length, or the length scale over which other physical processes occur (e.g. modal dispersion, four-wave mixing), the effect of disorder qualitatively changes the pulse propagation physics. In the case of modal dispersion, disordered coupling leads to a diffusive modal pulse broadening as the energy undergoes a random walk between different modes. In other words, instead of temporally broadening linearly with fiber length, as $\delta \beta^{p}_1 L$ as in Eqn. 8, pulses broaden instead in proportion to $\sqrt{L}$. The averaging, or wash-out, caused by disordered mode coupling can also reduce the cross-phase modulation and reduce or eliminate the four-wave mixing terms in \ref{eq3}\cite{Mecozzi2012,Mecozzi2012a,Mumtaz2013,Xiao2014, Mecozzi2016,Guasoni2017}. If the modes considered all occupy a single quasi-degenerate mode group and relevant length scales (the dispersion, nonlinear lengths) all exceed the correlation length, then the modes are in the so-called strong coupling limit of disorder. In this case, Eqn. 3 reduces to a multimode Manakov equation\cite{Mecozzi2012a}, for which exact soliton solutions exist. If multiple degenerate mode groups are present, a set of coupled Manakov equations\cite{Mecozzi2012} is obtained. On one hand, this elimination/reduction of four-wave mixing can be detrimental, such as for parametric amplification\cite{Guasoni2017}. On the other hand, for multimode fiber communications, strong disordered mode coupling leads a remarkable outcome: Due to nonlinear crosstalk averaging, for a given total power, the signal-to-noise ratio for a set of N randomly coupled modes is actually lower than for N uncoupled single-mode fibers\cite{Mecozzi2016}. 

For more details on modeling disorder within the MMNLSE framework, we refer readers to Refs. \cite{Guasoni2017,Mumtaz2013}, to Ref. \cite{Palmieri2014} for analytic expressions of disorder-induced mode-coupling coefficients for typical fiber manufacturing errors, and to Ref. \cite{Mecozzi2016} for propagation with the multimode Manakov and coupled Manakov equations.

\subsubsection{Saturating gain}
In multimode fibers with gain, the different modes not only receive gain, but necessarily compete for available inverted population since they overlap with the same gain medium. To lowest order, gain can be added into the GMMNLSE by an additional term $g_pA_p$, where $g_p$ is the small signal gain for the electric field in mode $p$. In most fiber amplifiers, however, saturation will be important. For ultrashort pulses, a common model for this is to let the gain depend on the time-integrated power, $g(I)=g_o/[1+\int \mathrm{d}t I(x,y,z,t)/I_{sat}]$. Incorporation of the full gain saturation will be described in a future publication. A first approximation to the saturation (which is often adequate) is to expand the saturating gain in a Taylor series, multiply on each side by a spatial eigenmode and integrate over space (i.e., perform a change of basis from the Cartesian coordinates to the spatial eigenmodes). This process yields the additional terms: $g_pA_p - \sum \limits_{l,m,n}^NS_{plmn}^RA_l\int_{-\infty}^{\infty} d\tau A_m(z,\tau)A_n^*(z,\tau)/I_{sat}$. 

\subsection{Important future directions}
If you were overwhelmed by the GMMNLSE, we hope that the tutorial above, and walking through the included examples, has helped to convert this bewilderment into an appreciation of the beautiful orchestral arrangements of multimode nonlinear optics. At this stage, however, we have to warn you of yet a second wave of whelming as we briefly describe important open directions. These directions include several that are relevant to relatively near-term applications in fiber lasers, telecommunications, and imaging, but there are also compelling directions for basic research. This echoes what has happened with the study of single-mode fiber, where applied science and engineering has supported, and also benefited from, research that uses single-mode fiber as an experimental test-bed for basic physics.
\subsubsection{Integrating disorder and nonlinearity}
For applications, understanding the role of disorder in the context of both linear and nonlinear multimode propagation will be critical. Disorder is not only inevitable in long fibers, but as mentioned above, strong linear mode coupling can have an almost fantastic benefit for telecommunications. That said, how and if this will actually be achieved is still unknown. The answer will need to consider not only the interaction between linear and nonlinear coupling effects, but also the practicalities and economics of a transition to multimode transmission\cite{Richardson2013}. For now, experimental demonstrations of multimode fibers with strong linear mode coupling (especially verifying the impact on nonlinearity) are severely lacking.  Several works that consider the complex intermediate coupling regime have interesting findings already\cite{Xiao2014,Wright2016,Guasoni2017,Rademacher2013}. Meanwhile, several powerful techniques are being used to characterize coupling within multimode fibers\cite{Nicholson2008,Nguyen2012,Carpenter2014,Ploschner2015,Carpenter2016,Xiong2016,Carpenter2017,Guang2017}. In coming years, these theoretical and experimental developments will need to come together. 

\subsubsection{Gain, amplifiers, and lasers}  
Very recently, several works have shown the potential for multimode propagation in fiber amplifiers and lasers to generate unprecedented high power, and even spatially- or spatiotemporally-engineered light\cite{Guenard2017,Florentin2016,Wright2017}. Before this, multimode fibers were considered as a means of implementing a fiber-format saturable absorber (e.g.,\cite{Winful1992,Proctor2005,Buttner2012,Nazemosadat2013}). The possibility for strong interactions between longitudinal modes of many different transverse eigenmode families means that mode-locked lasers with multimode fiber can display a stunning range of behaviors, including potentially orders-of-magnitude higher peak power than those based on single-mode, large-mode-area fibers. While these features are compelling, practical deployment of multimode fiber ultrafast oscillators is sure to be limited in the short term by complications of disorder, and from peak- and average-power related damage to the fiber medium. Multimode fiber amplifiers present a more near-term possibility to obtain spatially-engineered, high-power laser sources. The potential impact of just being able to harness the ultra-large mode area of modes in multimode fiber is significant\cite{Nicholson2015,Nicholson2016,Nicholson2017}. Further improvements will probably require an increasing attention to intermodal effects (as Example 3 suggests, for example). However, multimode fiber amplifiers like this already yield practical instruments with very compelling performance. They are likely to lead the way to a new generation of multimode fiber laser technologies.

A related direction concerns amplifiers for multimode fiber transmission systems. Nonlinear optical technologies may be particularly interesting here, as the unique capabilities of engineering the parametric gain suggest a route to ultrabroadband parametric amplifiers\cite{Guasoni2015}. If this can be accomplished, it could mean multimode fibers expand not only the spatial channel density of telecom lines, but also the number of spectral channels. Given that each spectral channel would represent $N$ more total channels (for each spatial degree of freedom), the impact of this technology could be substantial. 

\subsubsection{Wave turbulence and other fundamental nonlinear dynamics}  
The notion of turbulence in nonlinear optical wave propagation has produced many interesting results in the realm of quasi-1D systems and 2D systems\cite{Turitsyna2013,Picozzi2014,Aschieri2011,Kibler2011,Picozzi2008}. Pulse propagation in multimode waveguides can include a number of compelling ingredients beyond dispersion and nonlinearity, namely vortex formation in multiple different dimensional arrangements, different kinds of disorder\cite{Mafi2015,Mafi2017}, and dissipation. For this reason, we expect the study of optical wave turbulence in multimode fibers to find new phenomena, and to deepen the connection to fluid systems. Much like single-mode fibers have provided an unrivaled platform for studying nonlinear waves in one dimension, multimode fibers may provide a platform of equal utility for studying nonlinear dynamics in higher dimensions, and with controllable complexity. Just as single-mode fiber studies were enabled by the availability of cheap, high-quality fibers and measurement tools, scientific studies in multimode fiber will increasingly benefit from coming improvements in multimode fiber measurement tools, theoretical techniques, and other multimode technologies developed for applications like fiber lasers and telecommunications. It is difficult to estimate the long term impact of such fundamental science technologically. However, if single mode waveguides provide a historical template, then there is good reason to be excited about uncovering new physics, and reaching deeper understanding of important concepts within nonlinear multimode waveguides.  

\subsubsection{Optical signal processing and other nonlinear optical information technologies}  
As we move nearer to spatial division multiplexing, another area where multimode fibers may make a difference is in signal processing\cite{Hellwig2013,Hellwig2014,Schnack2015,Hellwig2016,Schnack2016}. Since intermodal interactions can be controlled with many more degrees of freedom than single mode fiber, and because spatial division multiplexing will require greater signal processing than single-mode transmission, inline signal processing, signal routing, etc. in multimode fibers could be important and see widespread use. More speculative applications of multimode fiber for computing are possible\cite{Wright2016,Mishra2017}. Quantum optics and information processing may also benefit from multimode waveguides\cite{Pourbeyram2016,Defienne2016,Cruz-Delgado2014}, including possibly high-dimensional entanglement with lower losses than most integrated platforms. 

\subsubsection{Multimode nonlinear optics beyond fused silica fibers}  
While we have exclusively considered multimode optical fibers here, there are many other platforms where we expect multimode nonlinear dynamics to occur, and where the treatment using the GMMNLSE will be valid. Recently, multimode effects are being explored in microresonators in the context of frequency combs\cite{Xue2015,Liu2014,DAguanno2016,Khurmi2016,Yang2016,Guo2017}. Multimode degrees of freedom allow qualitatively new kinds of solitons, which may allow for combs at different wavelength regimes, integrated dual comb functionality, compatibility with spatial-division multiplexing, with higher conversion efficiency, and so on. Bulk waveguides, or large stiff ``rod" fibers may make multimode fibers much more environmentally stable, and will allow highly multimode propagation with very large mode areas. While these devices will naturally sacrifice important practical features of fiber, such as flexibility or efficient heat dissipation, these downsides can be managed. If multimode fiber lasers and amplifiers relying on inflexible waveguides provide exceptional performance rivalling solid-state regenerative amplifiers, etc., the loss of these features will not diminish their attractiveness for applications compared to contemporary solutions. Finally, while multimode effects in hollow waveguides have been considered already\cite{Tani2013,Tani2014,Russell2014}, expanding these studies to highly multimode propagation may yield interesting results and practical benefits, such as ultrahigh power capacity, and increased flexibility of phase-matching the numerous unique effects possible in guided-wave gas-based nonlinear optics. 

Finally, there is much interest in generating supercontinuum in the infrared, using fibers comprised of chalcogenide glasses, for example. Generation of broad IR continua in few-mode fibers has been analyzed\cite{Ramsay2013,Kubat2016,Khalifa2017}, but experimental work has mostly focused on fundamental mode behavior\cite{Petersen2014} (a notable exception is Ref. \cite{Theberge2014}). Recent developments in GRIN multimode chalcogenide fibers\cite{Shabahang2017,Siwicki2017} may allow future continuum sources with higher power and more convenient pumping wavelengths. The numerous four-wave mixing processes in multimode fiber may allow broad supercontinuum generation to be seeded in the normal dispersion regime (typically below 4.5 $\mu$m in chalcogenide fibers). Since laser sources above 4.5 $\mu$m are rare and underdeveloped, this would be worthwhile.

\subsubsection{Engineering excitation, spatiotemporal and mode-resolved measurements}  
To date, most studies of nonlinear optics in multimode fibers have utilized relatively simple techniques to excite the fibers, and to make measurements. When interpreted correctly, simple measurements can be useful, especially for initial explorations. However, for studying specific phenomena and certainly for developing practical instruments, more sophisticated techniques are needed. More powerful techniques, such as spatial-light modulators, can not only allow better control of the field, but can also allow fascinating exploratory studies\cite{Florentin2016,Tzang2017}. Given the huge parameter space of a multimode fiber, we expect these kind of studies to produce surprising results. As multimode optics develops for telecommunications, the need for reliable spatial mode-resolved and spatiotemporal pulse measurement tools is increasing. Work along these lines has made impressive demonstrations\cite{Nicholson2008,Nguyen2012,Carpenter2014,Ploschner2015,Carpenter2016,Xiong2016,Carpenter2017,Guang2017,Pourbeyram2017}. If these techniques can be made accessible and widespread, they will significantly enhance scientific and technological developments in multimode waveguides and beyond. 

\section{Conclusion}
While in the beginning fiber optics emerged in the multimode domain, the development of low-loss single-mode fibers eventually underpinned the scientific and technological advances that have entirely transformed the internet (and the world). In the past few years, technological motivations have reinvigorated the study of multimode fibers. For nonlinear optics, this motivation led to theoretical developments, including notably the GMMNLSE, which we employ here as a tool to understand, both conceptually and numerically, multimode nonlinear wave propagation. The main features of multimode nonlinear wave propagation emerge from the different kinds of dispersion and the many different possible nonlinear interactions among modes. Linear mode coupling (often from disorder) and laser gain will likely increase in importance in the future. 

Just as the introduction of more and different instruments creates rich possibilities for composers, multiple modes add new spatial and temporal degrees of freedom to nonlinear optical wave propagation. As a result, qualitatively new optical capabilities and phenomena are being discovered. Applications include high-bandwidth telecommunications, imaging, and high-power laser sources. For the most part, these applications still require major developments before they become practical, but rapid progress is being made on a number of fronts. Fundamental scientific studies in multimode fibers may connect deeply to other topics throughout science, and may play the first notes of entirely new ideas ultimately paving the way for entirely new technologies.

\section*{Acknowledgment}
This work was supported by Office of Naval Research grant N00014-13-1-0649 and National Sciences Foundation grant ECCS-1609129. The  work of P.L.  was partially supported by the National Science Foundation  DMS-1412140. LGW thanks Walter Fu for helpful discussions. The codes described in this article are available for download online\cite{Code}. These codes include a freely-available mode solver from the MATLAB\textsuperscript{\textregistered} file exchange\cite{Fallahkhair2008}.

\ifCLASSOPTIONcaptionsoff
  \newpage
\fi



\bibliographystyle{IEEEtran}

%

	%
	%
	
	%
	
	\begin{IEEEbiographynophoto}{Logan G. Wright}
		received the B.S. degree in Engineering Physics from Queen's University, Kingston, Canada, in 2012. He is currently working toward the Ph.D. degree in applied physics at Cornell University, Ithaca, NY. His research interests include complex nonlinear optics and lasers.
	\end{IEEEbiographynophoto}
	
	\begin{IEEEbiographynophoto}{Zachary M. Ziegler}
		received the B.S. degree in Engineering Physics from Cornell University, Ithaca, NY in 2017. He is currently a research scientist at FeatureX in Boston, MA.
	\end{IEEEbiographynophoto}

	\begin{IEEEbiographynophoto}{Pavel M. Lushnikov}
 received the M.S. degree from Moscow Institute of Physics and Technology, Moscow, Russia in 1994, and the Ph.D.
degree from Landau Institute for Theoretical Physics, Moscow, Russia in 1997. He was the postdoctoral research at Landau Institute in 1998-1999 and at the Theoretical Division of Los Alamos National Laboratory in 1999-2003. In 2004-2006 he was Kenna Assistant Professor, Department of Mathematics, University of Notre Dame. He has been at the Faculty in the department of Mathematics and Statistics at University of New Mexico since 2006 as the Associate Professor (2006-2012) and the Full Professor since 2012 until now. He is also currently the visiting member of Landau Institute for Theoretical Physics and Visiting Scholar at the Theoretical Division of Los Alamos National Laboratory.
\end{IEEEbiographynophoto}	
	
	\begin{IEEEbiographynophoto}{Zimu Zhu}
	received his B.S. degree in Engineering Science specializing in Physics from the University of Toronto, Canada, in 2013. He is currently completing his Ph.D. degree at Cornell University, in Ithaca, NY. His research interests include the dynamics of nonlinear pulse propagation in multimode fibers.
\end{IEEEbiographynophoto}

	\begin{IEEEbiographynophoto}{M. Amin Eftekhar}
	 received the B.S. degree from  Shiraz University of Technology, Shiraz, Iran, and the M.S. degree from the University of Tehran, Tehran, Iran. He is currently working toward his Ph.D. degree in Optics and Photonicsat the College of Optics and Photonics, University of Central Florida. His research interests include nonlinear optics and fiber optics.
\end{IEEEbiographynophoto}	
	
	\begin{IEEEbiographynophoto}{Demetrios N. Christodoulides}	
	received the Ph.D. degree from The Johns Hopkins University, Baltimore, MD, USA, in 1986. He is the Cobb Family Endowed Chair and Pegasus Professor of Optics at the College of Optics and Photonics, University of Central Florida. He subsequently joined Bellcore as a Post-Doctoral Fellow at Murray Hill. Between 1988 and 2002, he was with the Faculty of the Department of Electrical Engineering, Lehigh University. His research interests include linear and nonlinear optical beam interactions, synthetic optical materials, optical solitons, and quantum electronics. His research initiated new innovation within the field, including the discovery of optical discrete solitons, Bragg and vector solitons in fibers, nonlinear surface waves, and the discovery of self-accelerating optical (Airy) beams. He has authored and coauthored more than 300 papers. He is a Fellow of the Optical Society of America and the American Physical Society. In 2011, he received the R.W. Wood Prize of OSA.
	\end{IEEEbiographynophoto}

	\begin{IEEEbiographynophoto}{Frank W. Wise}
		received the B.S. degree from Princeton  University, Princeton, NJ, the M.S. degree from the University of California, Berkeley, and the Ph.D. degree from Cornell University, Ithaca, NY. Since receiving the Ph.D. degree in 1988, he has been on the Faculty in the School of Engineering and Applied Physics at Cornell University.
	\end{IEEEbiographynophoto}	
	
	

\end{document}